\documentclass[fleqn,usenatbib]{mnras}

\usepackage{newtxtext,newtxmath}
\usepackage{ulem}

\usepackage[T1]{fontenc}

\DeclareRobustCommand{\VAN}[3]{#2}
\let\VANthebibliography\thebibliography
\def\thebibliography{\DeclareRobustCommand{\VAN}[3]{##3}\VANthebibliography}

\usepackage{graphicx}	
\usepackage{amsmath}	

\title[]{Plasma lensing interpretation of FRB 20201124A bursts at the end of September 2021}
\author[X. Chen et al.]{Xuechun Chen
,$^{1,2}$
 Bin Hu,$^{1,2}$\thanks{E-mail: bhu@bnu.edu.cn}
 Pei Wang,$^{3}$
 Wenwen Zheng,$^{4}$
 Di Li,$^{3}$
 Xinzhong Er$^{5}$
\\
$^{1}$Institute for Frontier in Astronomy and Astrophysics, Beijing Normal University, Beijing 102206, China\\
$^{2}$Department of Astronomy, Beijing Normal University, Beijing 100875, China\\
$^{3}$National Astronomical Observatories, Chinese Academy of Sciences, Beijing 100101, China\\
$^{4}$Purple Mountain Observatory, Chinese Academy of Sciences, Nanjing 210023, China\\
$^{5}$South-Western Institute for Astronomy Research, Yunnan University, Kunming 650500, China
}

\date{Accepted XXX. Received YYY; in original form ZZZ}

\begin{document}
\label{firstpage}
\pagerange{\pageref{firstpage}--\pageref{lastpage}}
\maketitle

\begin{abstract}
When the radio photons propagate through a non-uniform electron density volume, the plasma lensing effect can induce an extreme magnification to the observed flux at certain frequencies. Because the plasma lens acts as a diverging lens, it can extremely suppress the observed flux when aligned with source. These two properties can theoretically cause a highly magnified Fast Radio Burst (FRB) to faint or even disappear for a period of time. In this paper, we interpret that the significant increase in burst counts followed by a sudden quenching in FRB 20201124A in September 2021 can be attributed to plasma lensing. Based on the one-dimensional Gaussian lens model, we search for double main-peak structures in spectra just before its extinction on September 29, 2021. After the de-dispersion and de-scintillation procedures, we find eight bursts with double main-peaks at stable positions.
There are three parameters in our modelling, the height $N_0$ and width $a$ of the one-dimension Gaussian lens and its distance $D_{\mathrm{LS}}$ to the source. We reformulate them as a combined parameter $\mathrm{P}_0 \propto \left ( \frac{a}{\mathrm{AU}}\right )\sqrt{\frac{\mathrm{kpc}}{D_{\mathrm{LS}}} \frac{\mathrm{pc}\;\mathrm{cm}^{-3}}{N_0} }$. The frequency spectra can give an accurate estimation of $\mathrm{P}_0$ corresponding to $\left ( \frac{a}{\mathrm{AU}}\right )\sqrt{\frac{\mathrm{kpc}}{D_{\mathrm{LS}}} \frac{\mathrm{pc}\;\mathrm{cm}^{-3}}{N_0} } \approx 28.118$, while the time of arrival only give a relatively loose constraint on $a^2/D_{\mathrm{LS}}$.
Comparing with the observation dynamic spectra, we suggest that for a plasma lens in host galaxy, e.g., $D_{\mathrm{LS}}\approx 1\mathrm{kpc}$, the width of lens can not be larger than $40\mathrm{AU}$.
At last, we estimate the relative transverse motion velocity between the lens and source, $v\approx98\left(\frac{a}{\mathrm{AU}}\right)\mathrm{km/s}$.

\end{abstract}

\begin{keywords}
Radio transient sources: fast radio burst --- Gravitational lensing: strong --- Caustic crossing
\end{keywords}



\section{Introduction}
Past twenty years have seen the first discovery and rapid progressions of the mysterious pulses -- Fast Radio Bursts (FRBs) \citep{Lorimer2007,Cordes2019,Petroff2022}. Large excess of dispersion measure (DM) to the Milky Way's (MW) and locating host galaxy of some FRBs unveil their extragalactic origin \citep{Thornton2013,Chatterjee2017,Nimmo2022}. At the time of writing, over 600 FRBs have been discovered with $\sim20$ of them exhibiting multiple bursts \citep{2021ApJS..257...59C}. FRB population seems to be divided into two categories: repeating and one-off, though it is still under debate whether they actually belong to one category or not \citep{Spitler2016,Caleb2019}. Considerable amount of theoretical work has investigated FRB progenitors and most suggest an association with neutron star (NS) or magnetar \citep{Platts2019}. Promising models include the magnetar flare model\citep{Popov2010}, the model of NS-NS merger \citep{Totani2013,WangJS2016}, the collision model between neutron stars (highly magnetized pulsars) and asteroids/comets/asteroid belts \citep{Geng_2015,Dai2016,Dai2020}. The recent exciting discovery of FRB 200428 provides some indications that progenitors of FRBs (at least in part) are magnetars \citep{Bochenek2020,2020Natur.587...63L,2020Natur.587...54C}. Proposed radiation mechanisms consists of magnetospheric models \citep{Lu2020} and shock models \citep{Metzger2019}, which set the emitted radio signals at difference distances from the center star. We recommend Refs. \citep{ZhangBing2023,Xiao2021} for the review of the possible emission mechanism. 
Moreover, FRBs also exhibit some noteworthy phenomenas, such as narrow-band emission \citep{Gourdji2019}, frequency drifting\citep{Hessels2019}, quenching \citep{Zhoudejiang2022} in some repeaters and quasi-periodic behaviour in several one-off FRBs \citep{Chime2022}.

As a cosmological source, FRBs can be lensed by material along their propagation path to the earth and produce specific phenomenas, which make them a powerful probe. Repeating FRBs can be gravitational strong lensed by a galaxy appearing copies with fixed time series separated by several days, and this time lag can be used to constrain $H_0$ \citep{Li2018}. Under a strong lensed FRB system, measuring the differential birefringence of multiply images can be used to search axion, which is a main candidate of dark matter \citep{GaoRan2023}. 
A one-off FRB can be microlensed by an intermediate-mass black hole or a field of stars, exhibiting complex temporal profile \citep{Chen2021a,Chen2021b}. This provides an estimation for the lens mass. The radio wave can be also deflected by the inhomogeneous electron distribution, i.e., plasma lensing, which opens a new direction to study the ionized gas in the universe. A most characteristic of plasma lensing is the chromaticity. Frequency drifting or peak structures can appear in the spectrum of FRBs, offering constraints on the properties of the plasma structure \citep{Cordes2017,Hessels2019,Levkov2022}. Plasma lensing can also modify the gravitational lensing predictions, inducing changes in image positions, magnification and delays \citep{Er&Mao2014,Er2020,Er&Mao2022,Er2022a,Er2022b,Sun2022}. The missing central lensed image can be observed at low frequency if there is strong plasma lensing effect \citep{Er&Mao2022}. The large polarization variations of FRBs can be also attributed to the plasma lensing, and provides a potential probe to study the magnetic field in the universe \citep{Er2023}.
 Wave optic effects of the coherent radio sources are studied in Refs. \citep{Jow2021,Jow2020}. Lenses may locate near the source, half-way of the propagation path and in the MW. Studying these propagation effects can help distinct the intrinsic phenomena or detect the surrounding environment of source,  which is helpful to understand the burst mechanisms of FRBs.

FRB 20201124A is one of the most active repeating FRBs, which is first detected by CHIME\footnote{CHIME is acronym for Canadian Hydrogen Intensity Mapping Experiment \citep{Chime2018}} in Nov 2020. Subsequent bursts were detected in next months and a highly active phase came in April 2021 \citep{Chime2021,Lanman2022}. Many telescopes conducted follow-up observations to this source capturing more than one thousand bursts \citep{Main2021,Kumar2022,Xu2022}. During the 23-day (from September 25 to October 17) monitoring in 2021 by Five-hundred-meter Aperture Spherical radio Telescope (FAST, \citealt{Nan2011}), a quenching phenomenon was detected in FRB 20201124A \citep{Zhoudejiang2022}. There is an extraordinary active burst period from September 25 to 28. Within the total 4 hours observation time spanned during these 4 days, FAST detect 587 bursts with the integrated signal-to-noise ratio (SNR) > 7. And many of these bursts are extraordinary bright. Then, this source turned to an off-mode started from September 29, which last at least to October 17. Because of the observation schedule, FAST did not observe this source until February 2, 2022. And from March 20, we find this source becomes active again \citep{Zhoudejiang2022,Zhangyongkun2022,Jiangjinchen2022,Niujiarui2022}. Based on our description, we can conclude that the extinction of this source lasts at least 18 days. And it may last for half year.
Though another active repeater FRB 20121102 has also experienced a two-year quiescent period, we suggest that the event in FRB 20201124A is much distinctive and quite unusual. There is an exponential increase of the burst count in the first four days reaching an event rate of 380 per hour. However an abruptly extinguish was followed after such a sharp increase in burst count. We suspect that this feature can be attributed to the lensing modulation by a plasma structure along the propagation path of FRB \citep{Clegg1998,Cordes2017}. Flux is extremely suppressed when source is aligned with the lens \citep{Er2022b,Er&Mao2022}. As the lens, source, and observer move transversely, flux will be extremely magnified when crossing the lens caustic. Since plasma lensing is frequency-dependent the high amplification on caustic will manifest certain frequencies in the spectrum of FRB, which can help to verify and constrain the lens \citep{Levkov2022}. 

In this paper, we investigate the quenching event of FRB 20201124A from the view of plasma lensing caustic. We analyse the data before the sudden quenching and search for the common double high narrow peak structures among different bursts, a specific prediction from one-dimension Gaussian plasma lensing model, in spectrum. 
The paper organization is as follows: in Section \ref{sec:Theory} we introduce the basic theory of plasma lensing; in Section \ref{sec:Lens Model} we describe the one-dimension Gaussian lens model and the frequency dependence of the caustics; the data reduction are described in Section \ref{sec:Data Reduction}; the plasma lensing search is presented in Section \ref{sec:Analysis} and a summary is given in Section \ref{sec:Summary}.

\section{Theory}
\label{sec:Theory}
We begin with a general geometric description of light deflection caused by the plasma lens \citep{Schneider1992,Wagner2020}. For a light ray, there is a geometric relationship between positions in the image plane ($\pmb{\xi}$) and source plane ($\pmb{\eta}$)
\begin{eqnarray}
\label{eq:lens equation in physical coordinate}
\pmb{\eta}=\frac{D_{\mathrm{S}}}{D_{\mathrm{L}}}\pmb{\xi}-D_{\mathrm{LS}}\hat{\pmb{\alpha}}\left ( \pmb{\xi} \right ). 
\end{eqnarray}
$D_{\mathrm{LS}}$, $D_{\mathrm{S}}$, $D_{\mathrm{L}}$ is the angular diameter distance between the source and the lens, the source and the observer, the lens and the observer respectively. A geometric light ray propagates normal to the constant phase surface and refraction occurs when there is a large electron number density fluctuation (the plasma lens) on the path \citep{Born1980}. The refraction angle $\hat{\pmb{\alpha}}$ is related to the phase change $\phi_{\mathrm{lens}}$ as
\begin{eqnarray}
\label{eq:alpha hat}
\hat{\pmb{\alpha}}=-\frac{1}{k}\pmb{\nabla}\delta \phi_{\mathrm{lens}},
\end{eqnarray}
where $k$ is the wave number. For plasma lensing, $\delta \phi_{\mathrm{lens}}$ has the form of \begin{eqnarray}
\label{eq:delta phi}
\delta \phi_{\mathrm{lens}} \left ( \pmb{\xi} \right ) = -\lambda r_e N_e\left ( \pmb{\xi} \right ).
\end{eqnarray}
Here $\lambda$, $r_e$, $N_e$ is the wave length, the classical electron radius, the local electron column density, respectively. 
Eq.(\ref{eq:lens equation in physical coordinate}), (\ref{eq:alpha hat}) and (\ref{eq:delta phi}) gives the lens equation
\begin{eqnarray}
\label{eq:lens equation 1}
\pmb{\eta}=\frac{D_{\mathrm{S}}}{D_{\mathrm{L}}}\pmb{\xi}-\frac{c^2 r_e}{2 \pi \nu^2}D_{\mathrm{LS}}\pmb{\nabla}_{\xi}N_e\left ( \pmb{\xi} \right ) .
\end{eqnarray}
We rewrite Eq.(\ref{eq:lens equation 1}) in angular coordinate and perform a scaling, $\pmb{y}=\frac{D_{\mathrm{L}}}{D_{\mathrm{S}}}\frac{\pmb{\eta}}{a}, \; \pmb{x}=\frac{\pmb{\xi}}{a}$ to obtain the dimensionless form:
\begin{eqnarray}
\label{eq:the general lens equation}
\pmb{y}=\pmb{x}-\pmb{\nabla}\psi\left ( \pmb{x} \right ) = \pmb{x}-\alpha_l\pmb{\nabla}N\left(\pmb{x}\right).
\end{eqnarray}
$\psi \left(\pmb{x}\right)$ is the lensing potential, which is determined by the plasma lens properties (including $D_{\mathrm{LS}}$--the distance to source, $a$--the characteristic width of lens and $N_0$--the maximum electron column density), the observational frequency $\nu$ and the electron column density profile $N\left(\pmb{x}\right)$. It should be noted that $\pmb{\nabla}_{\xi}$ in Eq.(\ref{eq:lens equation 1}) takes gradient respect to $\pmb{\xi}$ and has the relation: $\pmb{\nabla}_{\xi}=\frac{1}{a}\pmb{\nabla}$ to the dimensionless angular gradient operator $\pmb{\nabla}$ in Eq.(\ref{eq:the general lens equation}). Here, we define a deflect angle factor 
\begin{eqnarray}
\label{eq:deflect angle factor}
\alpha_l=\frac{D_{\mathrm{L}}D_{\mathrm{LS}}}{D_{\mathrm{S}}}\frac{c^2 r_e N_0}{2 \pi a^2 \nu^2}=\frac{1}{\mathrm{P}_0^2 \nu^2}.
\end{eqnarray}
$\nu$ is the frequency in GHz and $\mathrm{P}_0$ is the combination of lens property parameters. For a plasma lens in the local environment of FRB (i.e., $D_{\mathrm{L}} \approx D_{\mathrm{S}}$)\footnote{Here we consider a plasma lens in the host galaxy. A similar relation $D_{\mathrm{LS}} \approx D_{\mathrm{S}}$ applies to a Galatic plasma lens.} $\mathrm{P}_0$ can be approximated as
\begin{eqnarray}
\mathrm{P}_0 &\approx& \sqrt{ \frac{2\pi a^2}{D_{\mathrm{LS}} c^2 r_e N_0}}\, \cdot \, 1\mathrm{GHz} \nonumber \\
\label{eq:P0 expression}
&\approx&  0.024148\times\left ( \frac{a}{\mathrm{AU}}\right ) \sqrt{\frac{\mathrm{kpc}}{D_{\mathrm{LS}}} \, \frac{\mathrm{pc \; cm^-3}}{N_0} } ,
\end{eqnarray}
 $\alpha_l^{-1/2}$ is a scaled frequency ($\nu$ is in GHz) with the relation of
\begin{eqnarray}
\alpha_l^{-1/2}=\mathrm{P}_0 \cdot \nu.
\end{eqnarray}
For convenience, this form will be used hereafter to describe the modulation of plasma lensing caustic on FRB spectrum. By analogy with gravitational lensing (GL), $\alpha_{l}$ works like the Einstein radius and frequency plays a role of mass. This scenario corresponds to a ``GL system'' where ``mass'' varies with frequency.

Solutions of Eq.(\ref{eq:the general lens equation}) give positions, flux gain and time of arrival (TOA) of lensed images. Numerical methods are needed to accomplish this in most cases. Due to geometric path change and refractive delay from the lens, TOA of an image at $\pmb{x}$ is
\begin{eqnarray}
\label{eq:time of arrival}
T\left(\pmb{x}, \pmb{y}\right) &=&\frac{a^2}{c}\frac{D_{\mathrm{S}}}{D_{\mathrm{L}} D_{\mathrm{LS}}}\left(1+z_{\mathrm{L}}\right)\left [ \frac{1}{2} \left ( \pmb{x}-\pmb{y} \right )^2-\psi \left(\pmb{x}\right) \right ].
\end{eqnarray}
Like in gravitational lensing, Jacobian matrix of lens mapping from $\pmb{x}$ to $\pmb{y}$ describes the local area distortion and the magnification (i.e., the gain) is given by
\begin{eqnarray}
\label{eq:magnification}
\mu = \left | \delta_{ij} - \alpha_l \frac{\partial^2 N \left ( \pmb{x} \right ) }{\partial x_{i} \partial x_{j}}  \right | ^{-1}\;
\end{eqnarray}
according to Eq.(\ref{eq:the general lens equation}). Subscripts $i$ and $j$ denote two dimensions in the lens plane. Critical/caustic curves are positions in the $\pmb{x}$-plane/$\pmb{y}$-plane, where the determinant of Jacobian matrix vanishes and magnification is infinite mathematically. Since Eq.(\ref{eq:magnification}) is also frequency dependent, there is extreme amplification for a certain frequency when the source locates at caustics, i.e., spikes in the spectrum will be expected. We will describe this effect detailly in the next section.

\section{Lens Model}
\label{sec:Lens Model}
\begin{figure}
    \centering
	\includegraphics[width=1.\columnwidth]{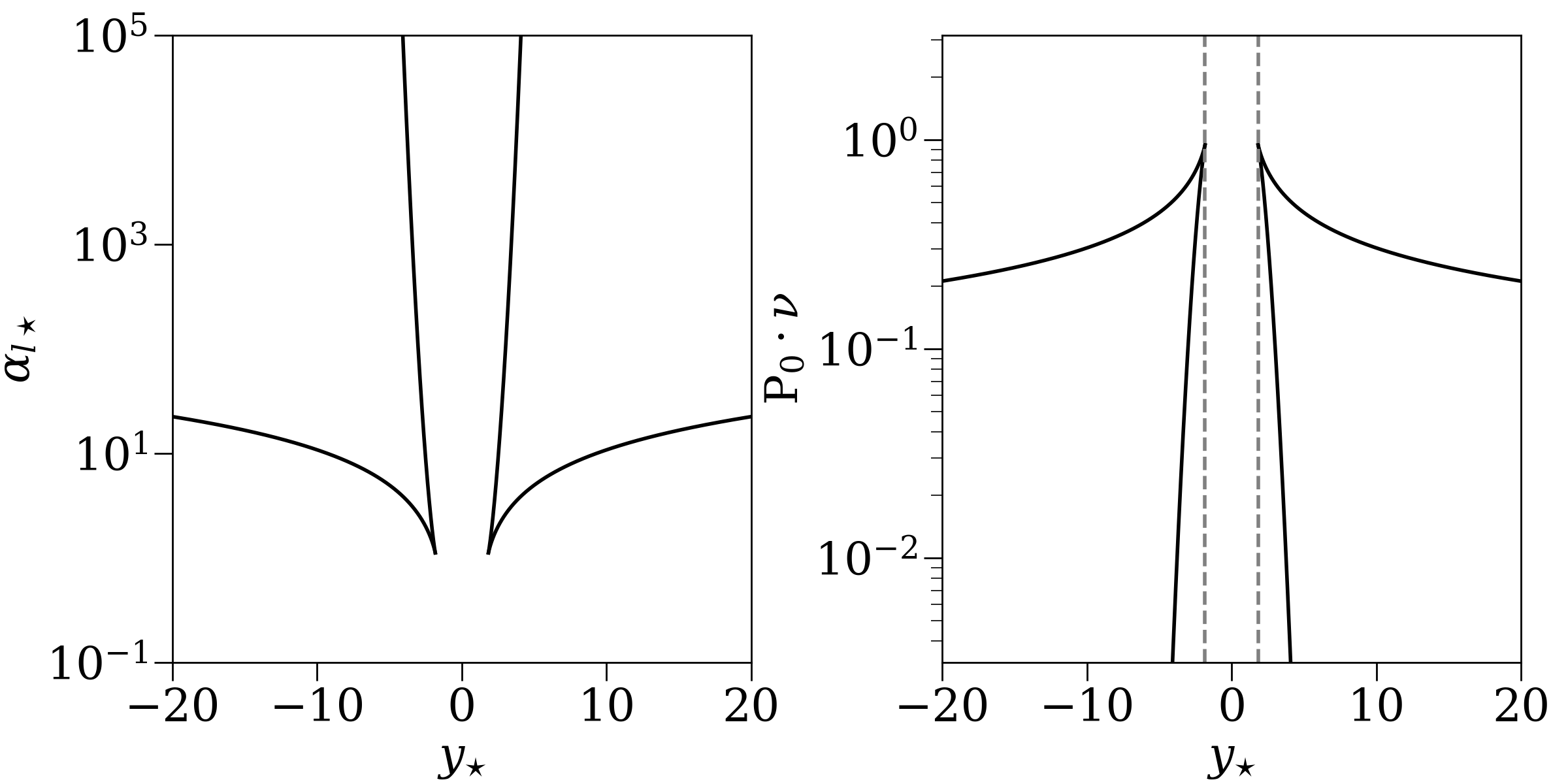}
    \caption{Relationship between the source position and the caustic frequency under one-dimensional Gaussian plasma lens model. $\mathit{Left}$: deflect angle factor $\alpha_{l}$ as a function of source position $y$ on caustic; $\mathit{Right}$: scaled frequency $\mathrm{P}_0 \cdot \nu$ as a function of source position $y$ on caustic.}
    \label{fig:P0_nu_vs_y_caustic}
\end{figure}

We use one-dimensional (1D) Gaussian profile to model the plasma macro lens \citep{Romani1987,Clegg1998, Cordes2017}, which may correspond to a long ionized filament structure in host galaxy in the magnetic environment. The 1D Gaussian model describes an extreme anisotropic lens, i.e., electron column density profile has the form of $N\left(\pmb{x}\right)=\mathrm{e}^{-x_0^2}$ and the deflection only occurs parallel to $x_0$ coordinate axis. Replacing $x_0$ by $x$ for convenience, the lens equation is written as 
\begin{eqnarray}
\label{eq:the initial dimensionless lens equation}
y=x+2\alpha_l x \mathrm{e}^{-x^2}.
\end{eqnarray}
According to Eq.(\ref{eq:magnification}) critical curve must satisfy
\begin{eqnarray}
\label{eq:alpha star vs x}
\alpha_{l\star}=\frac{\mathrm{e}^{x^2}}{4x^2-2}
\end{eqnarray}
and the lens equation on caustic is
\begin{eqnarray}
\label{eq:ystar vs x}
y_{\star}=\frac{2x^3}{2x^2-1},
\end{eqnarray}
where the subscript $\star$ denotes quantities on critical/caustic. For a given position of source the corresponding frequency with extremely magnification can be obtained by combining Eq.(\ref{eq:alpha star vs x}) and Eq.(\ref{eq:ystar vs x}), which is shown in Figure \ref{fig:P0_nu_vs_y_caustic}. 
For an electron column over density, $\alpha_l$ is always larger than zero and there will not be any critical curve for $x^2\le \frac{1}{2}$. We will also note that there is a theoretical minimum of $\left|y_{\star}\right|$ at $y_{\mathrm{min}}=\frac{3}{2}\sqrt{\frac{3}{2}}$ and the corresponding minimum value of $\alpha_{l\star}$ is $\alpha_{l\mathrm{min}}=\frac{\mathrm{e}^{\frac{3}{2}}}{4}$. As source position exceeds the minimum, it locates on caustics corresponding to two frequencies, i.e., watersheds of one-image and three-image systems, which induce extreme amplification to two frequency channels in the spectrum (given adequate resolution).

\begin{figure}
    \centering
	\includegraphics[width=1.\columnwidth]{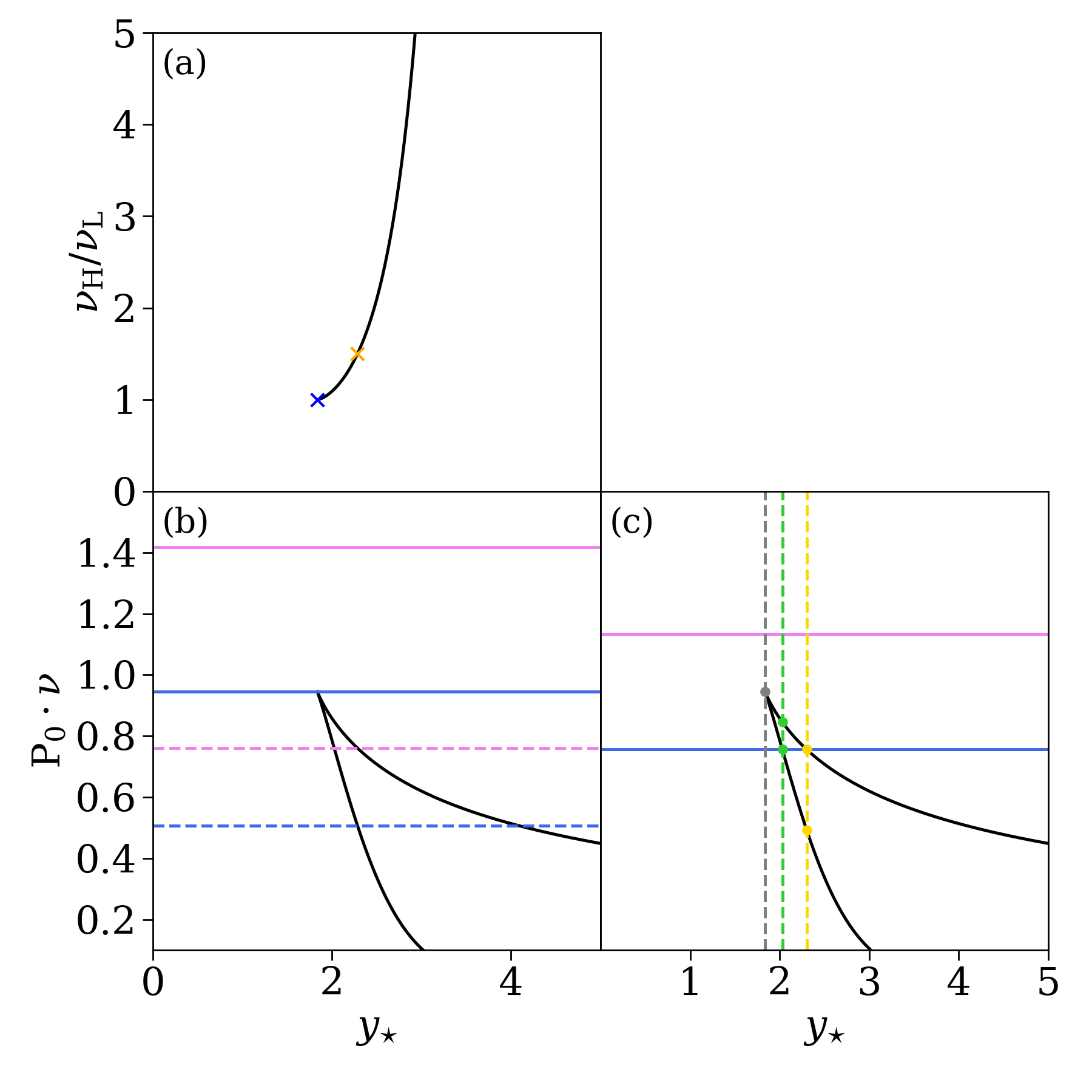}
    \caption{(a): Relationship between high/low frequency ratio and source position on caustic. Blue and orange crosses represent the position at $y_{\mathrm{min}}$ (i.e., $\nu_{\mathrm{H}}/\nu_{\mathrm{L}}=1$) and the position where $\nu_{\mathrm{H}}/\nu_{\mathrm{L}}=1.5$ respectively; (b): Situations where FAST cannot observe double peaks by plasma lensing caustic due to the limitation of observation bandwidth. If the observation window of FAST is as shown as the red-blue solid lines, the tip by caustics will not enter the window, and therefore no caustic frequency peak can be observed; If the observation window of FAST is as shown as the red-blue dashed lines, only single caustic peak can be observed; (c): An example that FAST can observe double peaks by plasma lensing caustic. The blue and violet horizontal solid lines represent observation bandwidth. The gray, green and yellow vertical dashed lines represent $y_{\mathrm{min}}$, the position where $\nu_{\mathrm{H}}/\nu_{\mathrm{L}}=1.5$ and the position where the theoretical high caustic frequency equals to the lower bound of observation bandwidth.}
    \label{fig:frequency_ratio_caustic}
\end{figure}

The deflect angle factor defined in Eq.(\ref{eq:deflect angle factor}) is a combination of radio frequency and lens properties. $\alpha_{l\star}$ is not a direct measurement in the spectrum of FRB because $\mathrm{P}_0$ is unknown. We therefore get the relation between the ratio of high frequency/low frequency and source position, which is shown in the sub-panel (a) of Figure \ref{fig:frequency_ratio_caustic}. From this, it can be seen whether a telescope, such as FAST, can observe double-peaked structure induced by lensing caustic mostly depends on the lower limit of observation bandwidth, i.e., where $\mathrm{P}_0 \cdot 1 \left(\mathrm{GHz}\right)$ locates in the right sub-panel of Figure \ref{fig:P0_nu_vs_y_caustic}. Therefore, FAST cannot see any double-peaked structure, if $\mathrm{P}_0 \cdot 1 \left(\mathrm{GHz}\right)$ is larger than the horizontal blue solid line in sub-panel (b) of Figure \ref{fig:frequency_ratio_caustic} or less than the horizontal blue dashed line. 
This provides a first limitation: $20<\left ( \frac{a}{\mathrm{AU}}\right ) \sqrt{\frac{\mathrm{kpc}}{D_{\mathrm{LS}}} \, \frac{\mathrm{pc \; cm^-3}}{N_0} }<40$ for FAST to capture the caustic double peaks according to Eq.(\ref{eq:P0 expression}). In the permitted region, FAST will see double peaks during the period defined by the vertical gray dashed line and the green dashed line. In addition, downward-drifting single peak will be seen during the period defined by the vertical green dashed line and gold dashed line given sufficient observation and bursts. Supposing a source approaches $y_{\mathrm{min}}$ from the right side and exceeds $y_{\mathrm{min}}$ towards the left side in sub-panel (c) of Figure \ref{fig:frequency_ratio_caustic}, it will be first extremely amplified, resulting in double peaks at fixed frequency among multiple bursts, and then suddenly enter a region of very low amplification. Since this scenario can qualitatively consist with the extinction behavior of FRB 20201124A at the end of Sep. 2021, we search for the plasma lensing hint, namely double peaks, in the data before extinction.

\section{Data Reduction}
\label{sec:Data Reduction}
We got the raw data (in PSRFITS\footnote{The introduction to PSRFITS can be referred to \cite{Hotan2004}} format) of FRB 20201124A on September 25th to 28th 2021, which can be referred to a series of papers \citep{Zhoudejiang2022, Zhangyongkun2022, Jiangjinchen2022, Niujiarui2022}. We obtain 587 bursts with the integrated SNR > 7. A number of propagation effects including dispersion, scintillation from the interstellar medium (ISM) can affect time-frequency signal of FRB. To obtain the final spectrum, we need to eliminate these effects. In this section we introduce the reduction procedures performed on the data in a nutshell. 

\subsection{Dedispersion}
The data span the [1000, 1500] MHz frequency range in 4096 channels. The arrival time of a radio signal with the frequency $\nu$ will be delayed by
\begin{eqnarray}
t_{\mathrm{d}}=4148.808\,\mathrm{s}\,\left ( \mathrm{DM/pc\;cm^{-3}} \right )\left ( \mathrm{MHz}/\nu \right )^2  
\end{eqnarray}
when propagating in cold plasma. $\mathrm{DM}$ is called dispersion measure, which is defined as the integral of free electron number density. We remove the dispersion in each frequency channel with DM value of $413.5\;\mathrm{pc\;cm^{-3}}$ from previous measurement \citep{Niujiarui2022}. Each FITS file contains data of 6.44s at the sampling time of $49.152\mu \mathrm{s}$. To reduce the file size, a down-sampling by a factor of 32 is performed to time and no down-sampling is performed to frequency. Therefore, we obtained the dedispersed data with a resolution of 0.122 MHz and 1.573 ms in frequency and time, respectively. Meanwhile, frequency channels with strong Radio Frequency Interference (RFI) are marked and replaced by the average of adjacent uncontaminated channels. More than 90\% channels pass the cut-and-fill processing. Following the definition of a burst in \citet{Zhoudejiang2022}, we cut a window of [TOA-50 ms, TOA+50 ms] in time axis from the de-dispersed data, i.e., 100 ms in total, which is used in next steps.

\subsection{Descintillation}
Small scale density fluctuations in turbulent plasma medium along the LOS will introduce diffractive interstellar scintillation (DISS), which also modulate the spectrum. Scintles are enhanced regions of flux density in dynamic spectrum and they usually have short time scale and narrow frequency bandwidth \citep{Lorimer2012}. The size in frequency is called scintillation bandwidth $\Delta \nu_{\mathrm{d}}$ (also referred to as de-correlation bandwidth). We use a 1D Gaussian kernel with the size $\sigma$ to perform convolution along the frequency axis of the dedispersed dynamic spectrum $S\left ( t,\nu \right ) $ as \citet{Levkov2022} did
\begin{eqnarray}
\label{eq:gaussian_smooth}
\bar{S}_{\sigma}\left ( t,\nu \right ) = \int \mathrm{d} \nu^{\prime} S\left ( t,\nu+\nu^{\prime} \right ) \frac{\mathrm{e}^{-\nu^{\prime 2}/2\sigma^2}}{\sqrt{2\pi}\sigma}.
\end{eqnarray}
This is to smooth the narrow-band scintillation structures and suppress the instrumental noise. Therefore, $\sigma$ is related to the scintillation bandwidth $\Delta \nu_{\mathrm{d}}$. The procedures we take to estimate $\Delta \nu_{\mathrm{d}}$ are based on auto-correlation function (ACF) method \citep{Cordes1986,Reardon2019,Wu2022} and mainly includes following steps. 
\begin{enumerate}
    \item We calculate the ACF of intrinsic spectrum $S\left(\nu\right)$ on the full 500MHz band as $\mathrm{ACF}^{\prime}\left(\Delta \nu \right)=\left<S\left(\nu\right)S\left(\nu+\Delta \nu\right)\right>$, where the $\left<\cdots\right>$ denotes for the integration over frequency $\nu$. 
    \item By fitting $\mathrm{ACF}^{\prime}\left(\Delta \nu\right)$ to a simple Gaussian profile $C_0 \mathrm{e}^{-C_1 \Delta \nu ^2}$ we get a quantifying size of intensity variation on a large frequency scale $f^{\prime}_{\mathrm{dc}}$\citep{Wu2022}.
    \item We construct a Gaussian kernel with $\sigma=6f^{\prime}_{\mathrm{dc}}$ to do a moving smooth on $S\left(\nu\right)$ and obtain $S_{\mathrm{smooth}}\left(\nu\right)$. 
    \item To get a precise estimation of scintillation on small frequency scale we need to do a similar thing in a suitable frequency region. We find the maximum of $S_{\mathrm{smooth}}\left(\nu\right)$ and define its left, right $3\sigma$ range as the interval $\left[\nu_1,\nu_2\right]$.
    \item We calculate the form $\delta S\left(\nu\right)=S\left(\nu\right)/S_{\mathrm{smooth}}\left(\nu\right)$ in $\left[\nu_1,\nu_2\right]$ and its $\mathrm{ACF}\left(\Delta \nu\right)=\left<\delta S\left(\nu\right)\delta S\left(\nu+\Delta \nu\right)\right>$.
    \item Finally we fit $\mathrm{ACF}\left(\Delta \nu\right)$ to model $\mathrm{ACF_{model}}\left(\Delta \nu \right)=\frac{m}{\Delta \nu_{\mathrm{d}}^2 + \Delta \nu^2}$ to get the scintillation bandwidth $\Delta \nu_{\mathrm{d}}$\citep{Masui2015}.
\end{enumerate}

\begin{figure}
    \centering
	\includegraphics[width=1.\columnwidth]{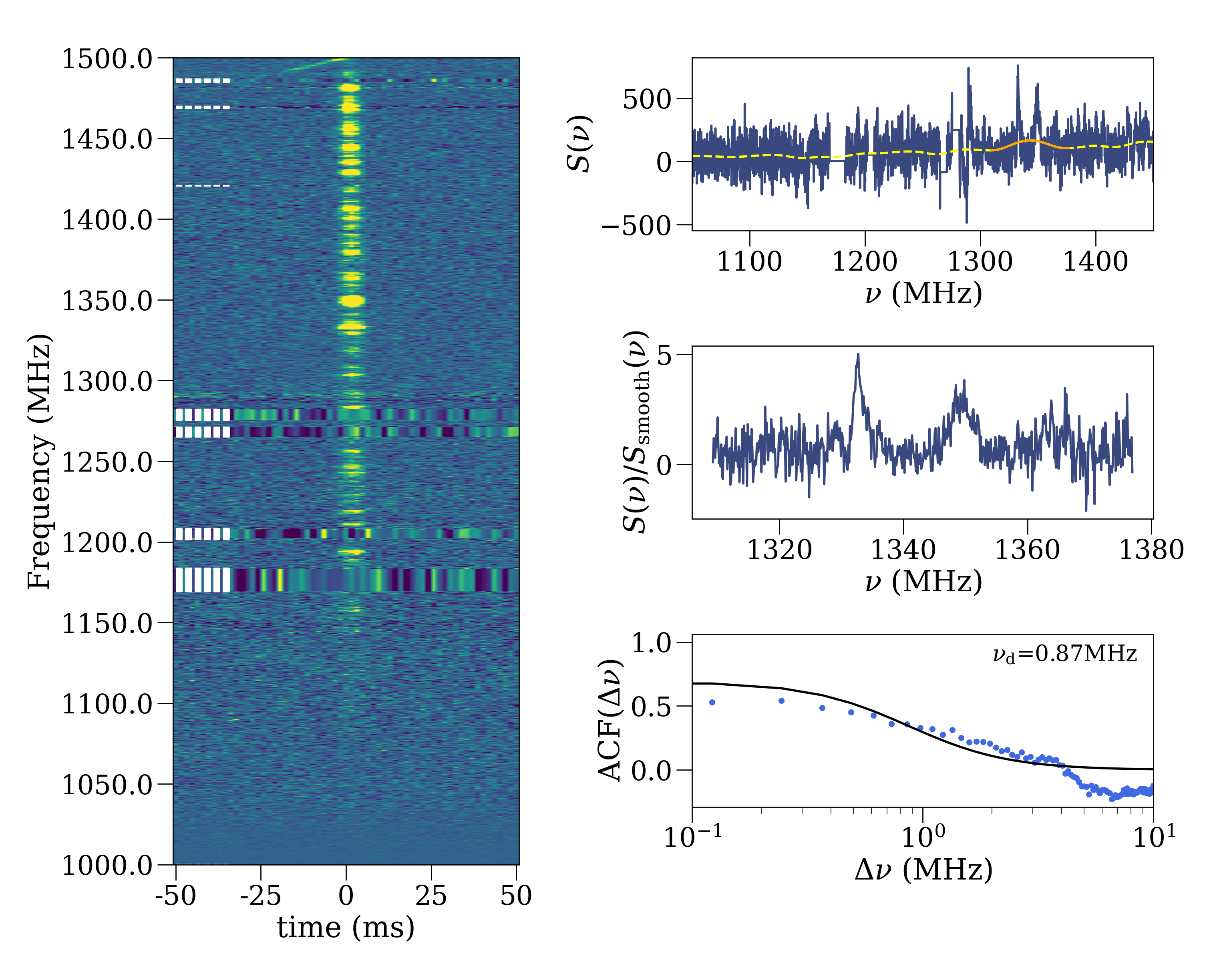}
    \caption{Calculating scintillation bandwidth from the intrinsic dynamic spectrum. $\mathit{Left}$: the intrinsic dynamic spectrum, white horizontal dashed lines mark the RFI channels; $\mathit{Right\;top}$: the dark blue solid line is the intrinsic spectrum $S\left(\nu\right)$, the yellow dashed line is the smooth spectrum $S_{\mathrm{smooth}}\left(\nu\right)$, the orange line denotes the frequency range $\left[\nu_1,\nu_2\right]$; $\mathit{Right \; middle}$: $\delta S \left(\nu\right)=S\left(\nu\right)/S_{\mathrm{smooth}}\left(\nu\right)$ calculated in $\left[\nu_1,\nu_2\right]$; $\mathit{Right \; bottom}$: the blue scatter is the ACF of spectrum in $\mathit{Right \; middle}$ panel and the black solid line is model fitting to ACF.}
    \label{fig:example_de_scintillation}
\end{figure}

\begin{figure}
    \centering
	\includegraphics[width=0.8\columnwidth]{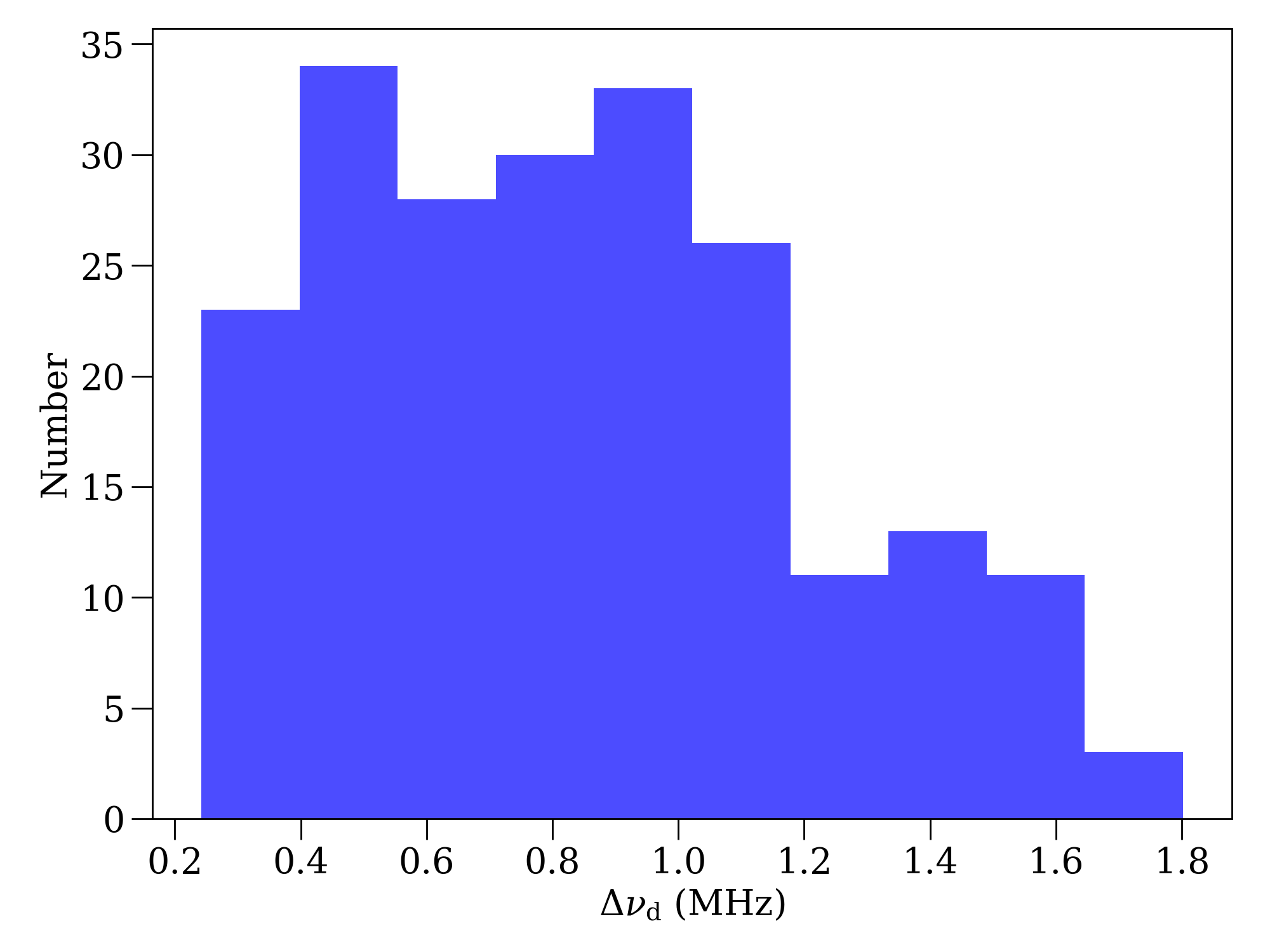}
    \caption{The scintillation bandwidth distribution of all bursts.}
    \label{fig:fdc_statistics}
\end{figure}

These procedures are summarized in Figure \ref{fig:example_de_scintillation}. The statistic histogram of $\Delta \nu_{\mathrm{d}}$ among all the bursts is shown in Figure \ref{fig:fdc_statistics} and the average of $\Delta \nu_{\mathrm{d}}$ is estimated to be $\sim 0.8\mathrm{MHz}$, which is also in agreement with the results of \citet{Main2022}, \citet{Zhoudejiang2022} etc.  
We adopted several times of $\Delta \nu_{\mathrm{d}}$ as a smooth window to construct the Gaussian kernel in Eq.(\ref{eq:gaussian_smooth}). Finally, we set $\sigma=\mathrm{4MHz}$ and the smoothed dynamic spectrum is as shown in the left sub-panel of Figure \ref{fig:smooth_dynamic_spectrum}. 

\subsection{Spectra}
\begin{figure}
    \centering
	\includegraphics[width=1.\columnwidth]{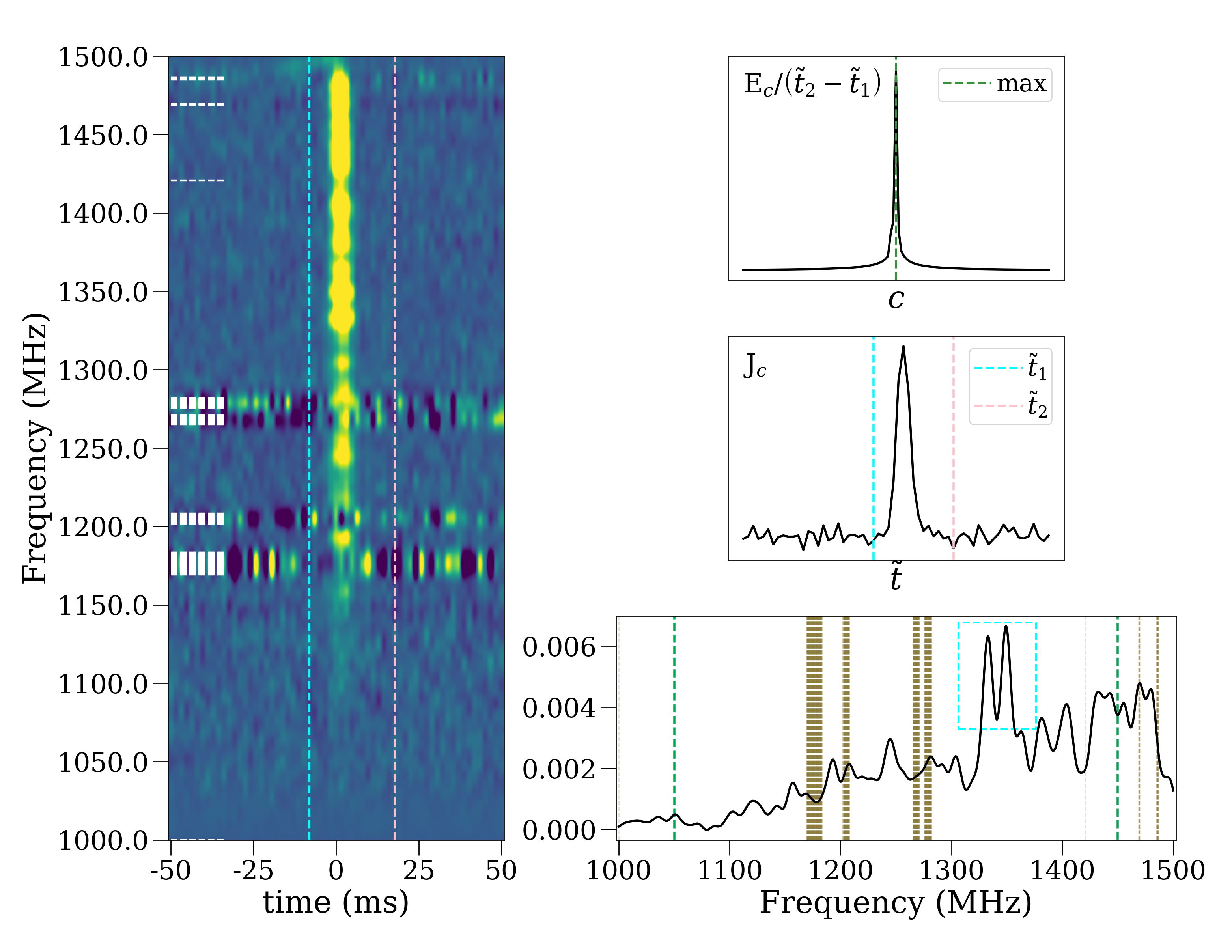}
    \caption{Obtaining spectrum from the smoothed dynamic spectrum. $\mathit{Left}$: the intrinsic dynamic spectrum in left sub-panel of Figure \ref{fig:example_de_scintillation} convolved by the Gaussian kernel defined by Eq.(\ref{eq:gaussian_smooth}). The blue and pink dashed lines denote the integration region boundary; $\mathit{Right\;top}$: a supplementary quantity $\mathrm{E}_c$ with the same meaning as \citet{Levkov2022}, which helps to determine the slope of integration region boundary; $\mathit{Right\;middle}$: a supplementary quantity $\mathrm{J}_c$ with the same meaning as \citet{Levkov2022}, which helps to determine the intercept of integration region boundary at time-axis; $\mathit{Right\;bottom}$: spectrum obtained from the integration region of smoothed dynamic spectrum.}
    \label{fig:smooth_dynamic_spectrum}
\end{figure}

After a smoothing procedure performed to the intrinsic dynamic spectrum, we next define an integral region over which flux density is projected towards frequency axis. The integral region is the area surrounded by the blue and pink dashed line in the left sub-panel of Figure \ref{fig:smooth_dynamic_spectrum}. The method is referred to \citet{Levkov2022}, which can help to maximize the spectral information. The final spectrum (normalized to one) obtained is shown in the right bottom sub-panel of Figure \ref{fig:smooth_dynamic_spectrum}. From this figure, we can also notice that there are two dominating peaks between 1300MHz and 1400MHz in the spectrum (as highlighted in the cyan dashed box). This spectral feature is very similar to that found in \citet{Levkov2022}. Detailed descriptions about these spectra including related structural features will be presented in the next section.

\section{Plasma lensing analysis}
\label{sec:Analysis}

\subsection{Searching for Double Peaks}
In the following analysis, 292 bursts with SNR higher than 50 are selected. With the goal of searching for double main peaks, we perform a \texttt{find\_peaks}\footnote{\url{https://docs.scipy.org/doc/scipy/reference/generated/scipy.signal.find_peaks.html}} procedure over the full-band spectrum of each burst. Here, term ``main peak" used in \citep{Levkov2022} is continued in this paper to describe a peak structure modulated by plasma lensing caustic. For the definition of a main peak,  we consider both its absolute height and relative height. We arranged the identified peaks in decreasing order of their heights. Two arrays $H_1$ and $H_2$ are set in the following way: $H_1\left[i\right]$ records the ratio of the $i$-th peak to the first peak (maximum) and $H_2\left[i\right]$ records the ratio of the $i$-th peak to the $\left(i-1\right)$-th. For the $i$-th peak, if $H_1\left[i\right]>70\%$ and  $H_2\left[i\right]>70\%$ we define it as a main peak. Finally, 45 bursts (including 23 on 09/28, 13 on 09/27, 5 on 09/26 and 4 on 09/25) are identified to have two main peaks in the full-band spectrum. 

\begin{figure}
    \centering
    \includegraphics[width=1.\columnwidth]{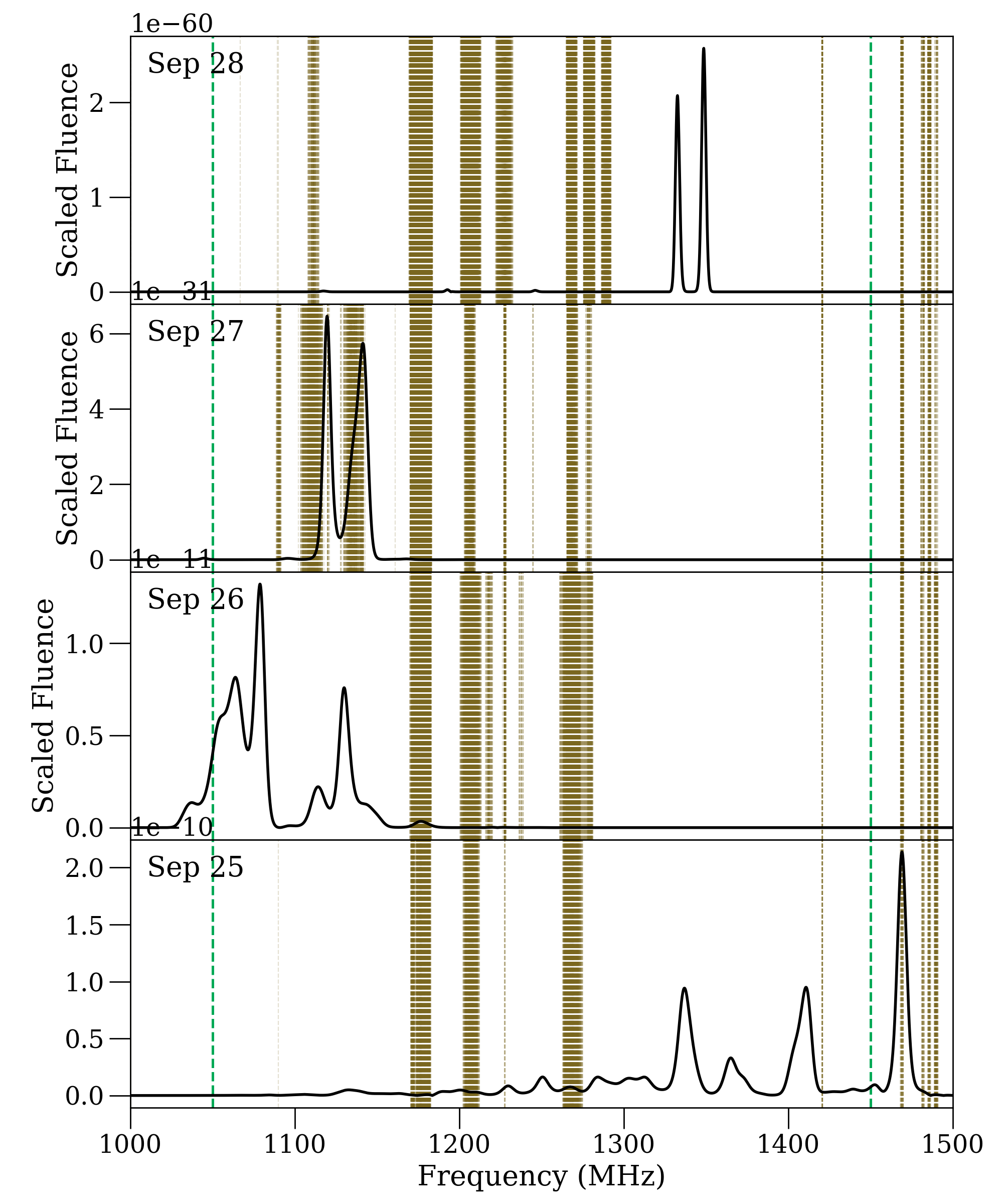}
    \caption{Multiplied normalized spectra of bursts in which only two main peaks are identified. Sub-panels from top to bottom represent the four days just before FRB 20201124A quenched.}
    \label{fig:define_main_peak}
\end{figure}

To see whether there is any common feature in bursts with double main peaks, we multiply together the normalized spectra belong to the same days. We show the results in Figure \ref{fig:define_main_peak} on 28th $\sim$ 25th from top to bottom.
 We notice that double-main-peak candidates found on the day just before the suddenly quenching (28th Sep) have a stable signature between 1300MHz and 1400MHz, which also locate far away from RFI channels. Eight of the 23 candidates have this stable feature, i.e., their double main peaks locate in [1300,1400]MHz. Since lensing modulation to spectrum should be relatively stable, these eight bursts are regarded as high-confident lensing candidates for next analysis (called  golden sample hereafter). Double-main-peak bursts on 25th and 26th show a complex profile after multiplication. It reflects that the double main peaks on these two days are unstable. Double main peaks are in RFI channels for bursts on 27th. Meanwhile these three days are not close to the suddenly extinction. Therefore our next analysis is based on the eight golden sample bursts on 28th. 

\begin{figure}
    \centering
    \includegraphics[width=1.\columnwidth]{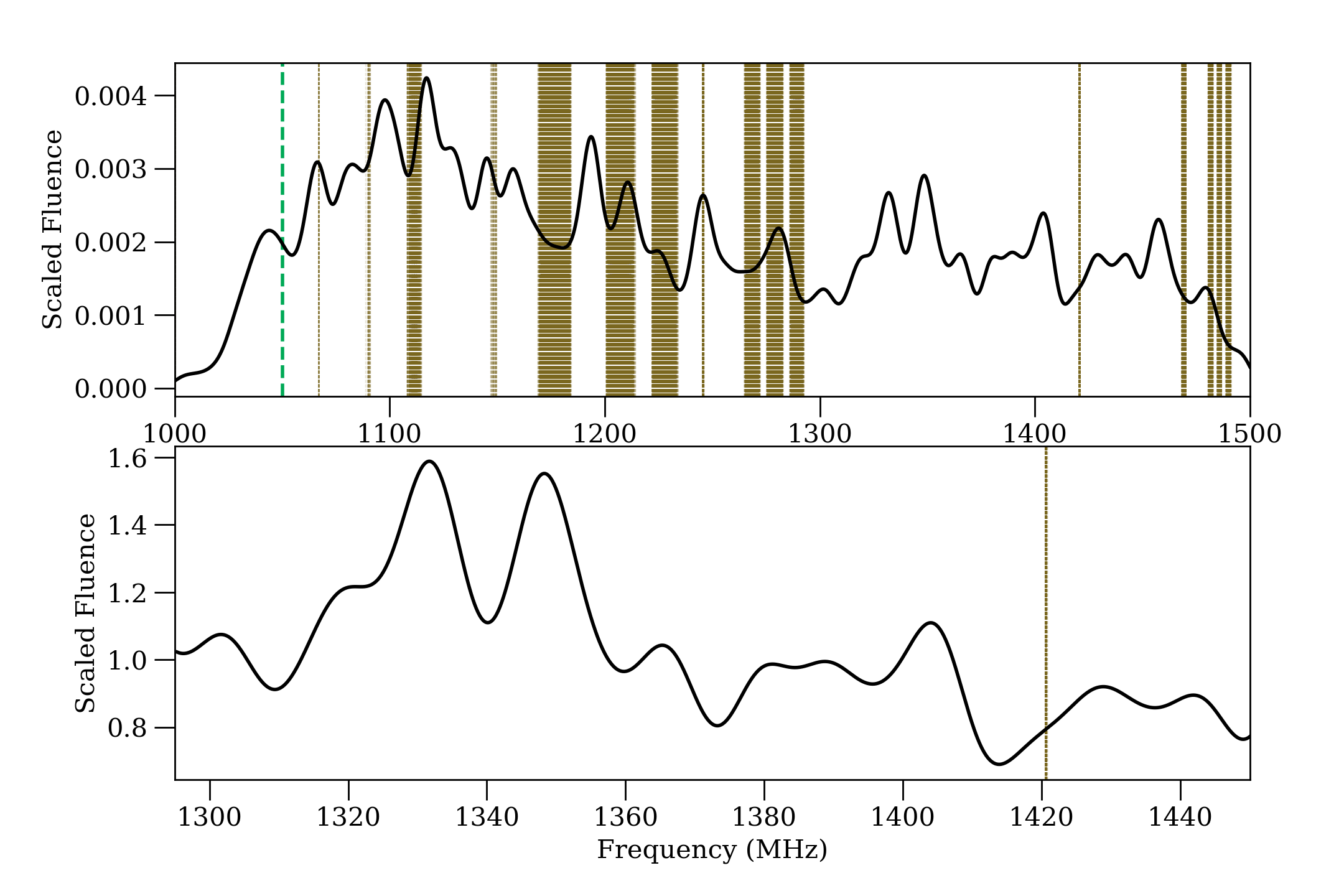}
    \caption{Accumulated normalized spectra of bursts on the day just before FRB 20201124A quenched, namely Sep 28. The upper panel shows the wide range frequency spectrum; the lower panel highlight the range between 1300 to 1440 MHz.}
    \label{fig:overal_spectrum}
\end{figure}

Furthermore, we accumulate all the normalized spectra of bursts on 28th due to concerns about the selection effect. They are plotted in the upper sub-panel of Figure \ref{fig:overal_spectrum}. There is a relative ``clean'' interval [1295, 1420]MHz, which is far away from all the RFI channels. We zoom in the results in this interval. It indicates that stable double main peaks in [1300,1400]MHz may exist in spectra of many bursts even if not all, which is qualitatively consistent with the caustic expectation. Therefore, we are more inclined to suppose that double-main-peaks in [1300,1400]MHz on Sep 28 is a possible plasma lensing caustic modulation. Under this assumption, we will now demonstrate to what extent we can extract the lens property from the data. 

\subsection{Lens Reconstruction}

\begin{figure}
    \centering
    \includegraphics[width=1\columnwidth]{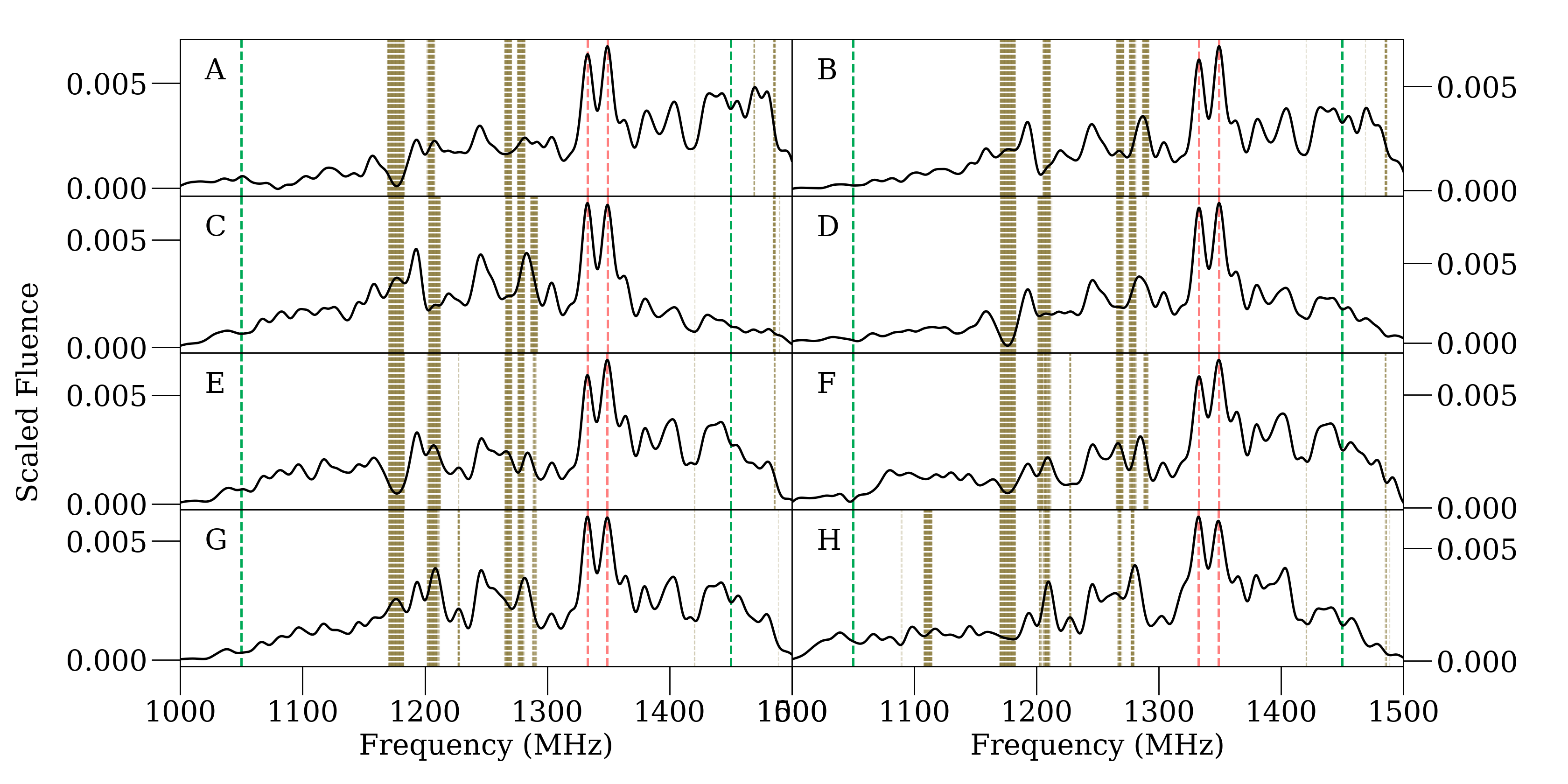}
    \caption{Spectra of eight bursts where the double-peak structure is suspected to be induced by plasma lensing caustic, i.e., the golden sample. Red dashed lines mark the main peaks. Yellow dashed lines denote the frequency channel with RFI. Green dashed lines correspond to 1050MHz and 1450MHz that are near the limit of observation bandwidth.}
    \label{fig:eight_level_1}
\end{figure}
\begin{figure}
    \centering
    \includegraphics[width=1.\columnwidth]{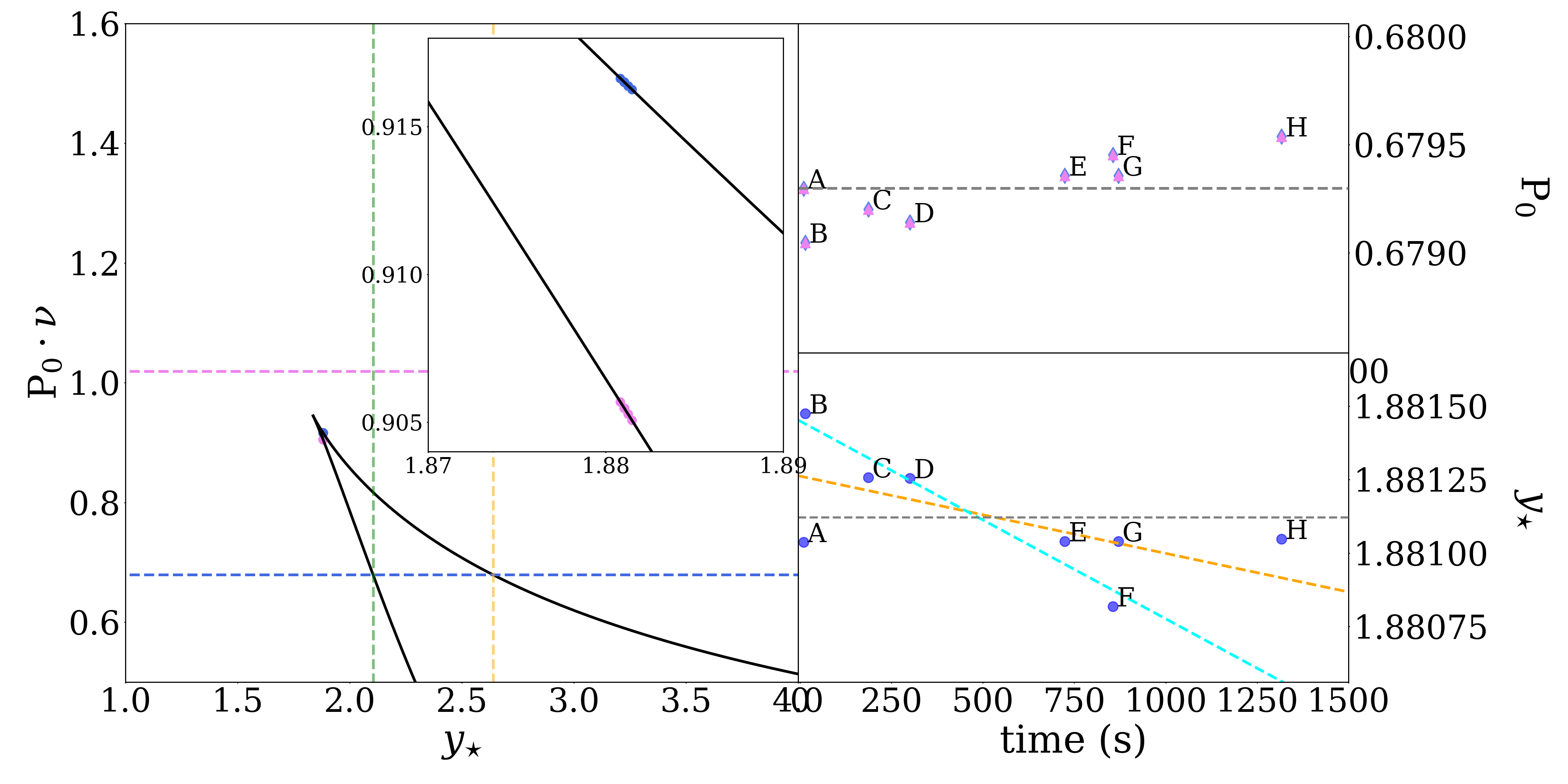}
    \caption{Fitting results of one-dimensional Gaussian model using double peaks in spectra of eight golden sample bursts. $\mathit{Left}$: The relationship between the scaled frequency $\mathrm{P}_0 \cdot \nu$ and the source position $y_{\star}$ on caustic. Blue and violet dots in zoom-in sub-panel represent source positions and corresponding caustic frequencies estimated by the frequency ratio of eight golden sample bursts. $\mathit{Top\,Right}$: $\mathrm{P}_0$ estimated for the eight golden sample bursts. The blue diamond and violet triangle denote for the $\mathrm{P}_0$ estimation by using lower and higher peak frequency, respectively. $\mathit{Bottom\,Right}$: Relation between burst time and source position estimated by frequency ratio.}
    \label{fig:Fit_lens}
\end{figure}

We show eight bursts with double main peaks induced by the possible lensing caustic, i.e., the golden sample in Figure \ref{fig:eight_level_1}. Peaks almost locate at the same position in spectra of different bursts. 
As is shown in the right sub-panel of Figure \ref{fig:P0_nu_vs_y_caustic}, with the ratio $q$ information of high frequency versus low frequency of each burst, we can obtain the theoretical source position $y$ and the corresponding scaled frequency $\mathrm{P}_0 \cdot \nu$. The result is illustrated in Figure \ref{fig:Fit_lens}. 
Through the eight bursts we estimate that the frequency ratio $q$ is in range 1.01209$\sim$1.01237 and the source position $y_{\star}$ is in range 1.88082$\sim$1.88147. $\mathrm{P}_{0}$ calculated from high frequency and low frequency is almost the same with the mean value of 0.67930 and scatter of $1.5\times 10^{-4}$ as is shown in the top right panel of Figure \ref{fig:Fit_lens}. This indicates that if the double main peaks of spectrum in these eight bursts are caused by plasma lensing caustic, they share the same lens properties as
\begin{eqnarray}
\label{eq:the combined form of lens parameters}
 \left ( \frac{a}{\mathrm{AU}}\right )\sqrt{\frac{\mathrm{kpc}}{D_{\mathrm{LS}}} \frac{\mathrm{pc}\;\mathrm{cm}^{-3}}{N_0} } \approx 28.118,  
\end{eqnarray}
which is a necessary condition for the plasma lensing assumption. 

Next with the estimated $\mathrm{P}_0$ value, we can find the according observation frequency range for FAST, which is shown by the blue and violet horizontal dashed lines in left panel of Figure \ref{fig:Fit_lens}. $y$ between the cusp connected by two black solid lines at $y_{\mathrm{min}}$ and the right green vertical dashed line at $y_{2}$ make up a region where FAST can observe double main peaks theoretically, i.e., [$y_{\mathrm{min}}$, $y_{2}$], where $y_{\mathrm{min}}=1.837,\; y_{2}=2.104$. Between the green and orange vertical dashed lines is the interval [$y_{2}$, $y_{1}$], where $y_{1}=2.639$. In this interval FAST can theoretically observe only one main peak due to bandwidth limitation and the right side of orange line at $y_{1}$ is the interval where no main peak can be observed by FAST.

We then perform a linear-fitting between the theoretical source position $y$ and burst time to the eight bursts in the right bottom panel of Figure \ref{fig:Fit_lens}, i.e., the orange dashed line. The transverse relative motion velocity between the lens, source and observer is therefore estimated to be $v_1\approx38\left(\frac{a}{\mathrm{AU}}\right)\mathrm{km/s}$; Considering that the transverse motion may be not strictly linear and monotonic, we use five bursts whose theoretical source positions monotonically change to do the same fitting (i.e., the cyan dashed line) and get the result of $v_2\approx98\left(\frac{a}{\mathrm{AU}}\right)\mathrm{km/s}$. For the estimation to transverse relative velocity, we also take another consideration. Since we did not find any double main-peak structures in the data before the second last day of quenching, we suspect that the source may move outside of the main-peak region (both double and single) for FAST observation within the time of one day. By this way, a minimum velocity is estimated by $\left.v_3> \left(y_{1}-y_{\star}\right) \middle / 1\mathrm{day} \right. \approx 1272 \left(\frac{a}{\mathrm{AU}}\right)\mathrm{km/s}$. As is shown in Figure \ref{fig:P0_nu_vs_y_caustic}, the region between two tips connected by black solid lines provides an estimation for the quenched time, whose distance is $2y_{\mathrm{min}}$. Since no burst was detected until Oct 17, we can estimate the relative speed as $\left.v_4< \left(y_{\mathrm{cr}}+y_{\star}\right) \middle / 19\mathrm{day} \right. \approx 339 \left(\frac{a}{\mathrm{AU}}\right)\mathrm{km/s}$ and $\left.v_5> \left(y_{\star}-y_{\mathrm{cr}}\right) \middle / 1\mathrm{day} \right. \approx 76 \left(\frac{a}{\mathrm{AU}}\right)\mathrm{km/s}$.
The estimations for $v_1$ to $v_5$ are summarized as follows:
 \begin{itemize}
     \item $v_1\approx38\left(\frac{a}{\mathrm{AU}}\right)\mathrm{km/s}$
     \item $v_2\approx98\left(\frac{a}{\mathrm{AU}}\right)\mathrm{km/s}$
     \item $v_3>1272 \left(\frac{a}{\mathrm{AU}}\right)\mathrm{km/s}$
     \item $v_4<339 \left(\frac{a}{\mathrm{AU}}\right)\mathrm{km/s}$
     \item $v_5>76 \left(\frac{a}{\mathrm{AU}}\right)\mathrm{km/s}$
 \end{itemize}
Taken $v_1 \sim v_5$ together we suggest the speed $\sim98\left(\frac{a}{\mathrm{AU}}\right)\mathrm{km/s}$ most reasonable and we are therefore more inclined to believe that the reason why we did not find significant double main peaks on Sep 25 $\sim$ 27 is due to inherent radiation mechanism.

\begin{figure}
    \centering
    \includegraphics[width=1.\columnwidth]{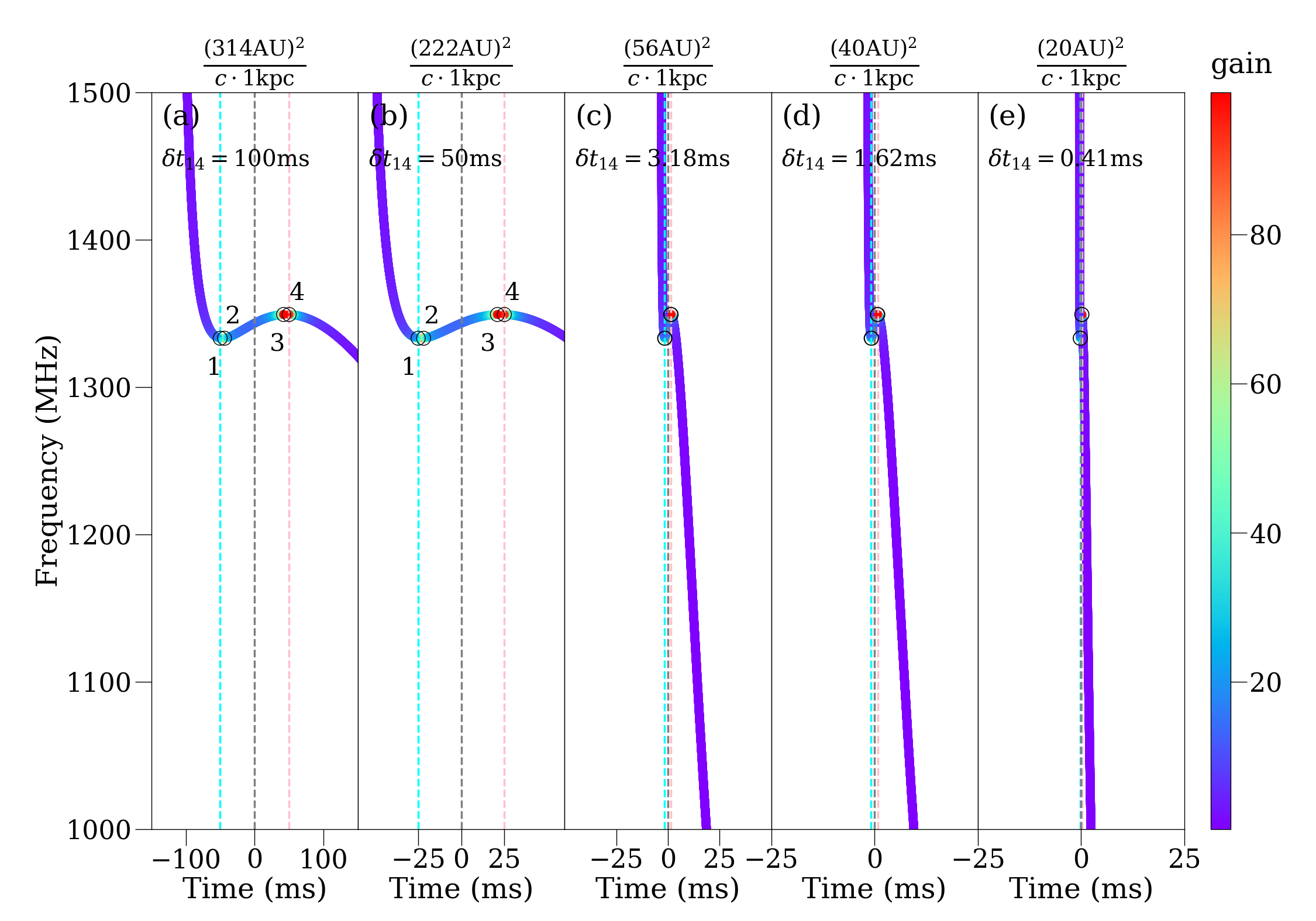}
    \caption{Solved images in 4096 frequency channels. Color of scatters denote the gain of each image. The four open black circles represent two pairs of counter-images in the caustic channels. Sub-panels (a)$\sim$(e) show time scales of $\delta t_{14}$ under different $\frac{a^2}{c \cdot D_{\mathrm{LS}}}$.}
    \label{fig:check_dynamic_spectra_2}
\end{figure}

Note that $y$ is a dimensionless length quantity scaled by the characteristic width $a$ of the lens. Next, we will consider inferring this width based on temporal information in the dynamic spectra of FRB. We perform an image search algorithm\footnote{The image search algorithm is described in \cite{Chen2021b}.} for the average $y=1.88112$ and $\mathrm{P}_{0}=0.6793$ through 4096 frequency channels. This actually solves 4096 lens equations in Eq.(\ref{eq:the initial dimensionless lens equation}) with varying $\alpha_{l}$. In this way, the number of images, the dimensionless image positions, magnification and dimensionless TOA have been fixed in each channel. However, the time domain is not fixed since the dimensionless TOA should be converted to the physical form by a factor of $\sim \frac{a^2}{c \cdot D_{\mathrm{LS}}}$. In Figure \ref{fig:check_dynamic_spectra_2}, we plot the solved images in each frequency channel. We find channels where caustic frequencies belong to. Two pairs \footnote{The green and red spots highlighted in Figure \ref{fig:check_dynamic_spectra_2} belongs to two three-image systems, in which one of the images is moderately magnified and two are highly magnified. The pairs we mentioned in the main text are the highly magnified images.} of counter-images with extremely high magnification are formed in these two frequency channels, corresponding to points ``1'', ``2'', ``3'' and ``4'' respectively in Figure \ref{fig:check_dynamic_spectra_2}. We define $\delta t_{14}$ as the time delay between the first arrival image and the last one. For different $\frac{a^2}{c \cdot D_{\mathrm{LS}}}$ the time scale of $\delta t_{14}$ is significantly different. Sub-panels in Figure \ref{fig:check_dynamic_spectra_2} show five examples:
\begin{itemize}
    \item (a)--$\delta t_{14}$=100ms. Due to the limitation of time window, the dynamic spectrum cannot capture all the high-magnification images if $\delta t_{14}>$100ms. Therefore the spectrum will not show double main peaks.
    \item (b)--$\delta t_{14}$=50ms. Actually due to the usual definition of a burst, it is also unable to capture the double main peaks in this case.
    \item (c)--$\delta t_{14}$=3.18ms. In this case $\delta t_{14}$ is twice the time resolution of dynamic spectrum. Above this value the dynamic spectrum exhibits obvious separate components.
    \item (d)--$\delta t_{14}$=1.62ms. In this case $\delta t_{14}$ is just the same as the time resolution of the dynamic spectrum.
    \item (e)--$\delta t_{14}$=0.41ms. In this case $\delta t_{14}$ is equal to the standard deviation of a Gaussian profile with the FWHM of 1ms.
\end{itemize}

\begin{figure}
    \centering
    \includegraphics[width=1.\columnwidth]{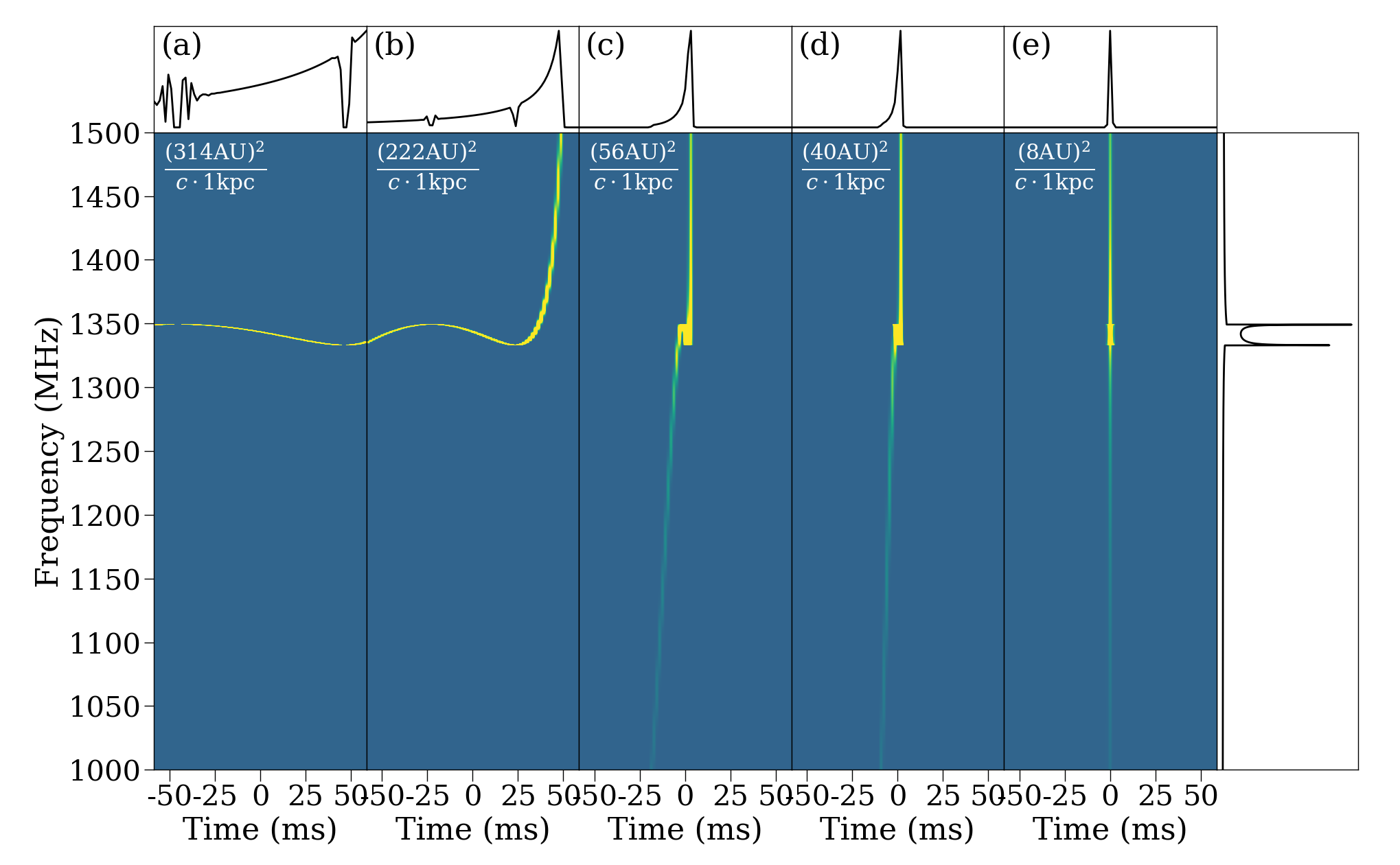}
    \caption{Simulated dynamic spectra for cases in Figure \ref{fig:check_dynamic_spectra_2}. The time axis spans 120ms in each sub-panel. The spectra of all the five cases are the same, as shown in the right sub-panel. The five upper sub-panels show the time-domain light curves.}
    \label{fig:plot_dynamic_spectra}
\end{figure}

\begin{figure}
    \centering
    \includegraphics[width=1.\columnwidth]{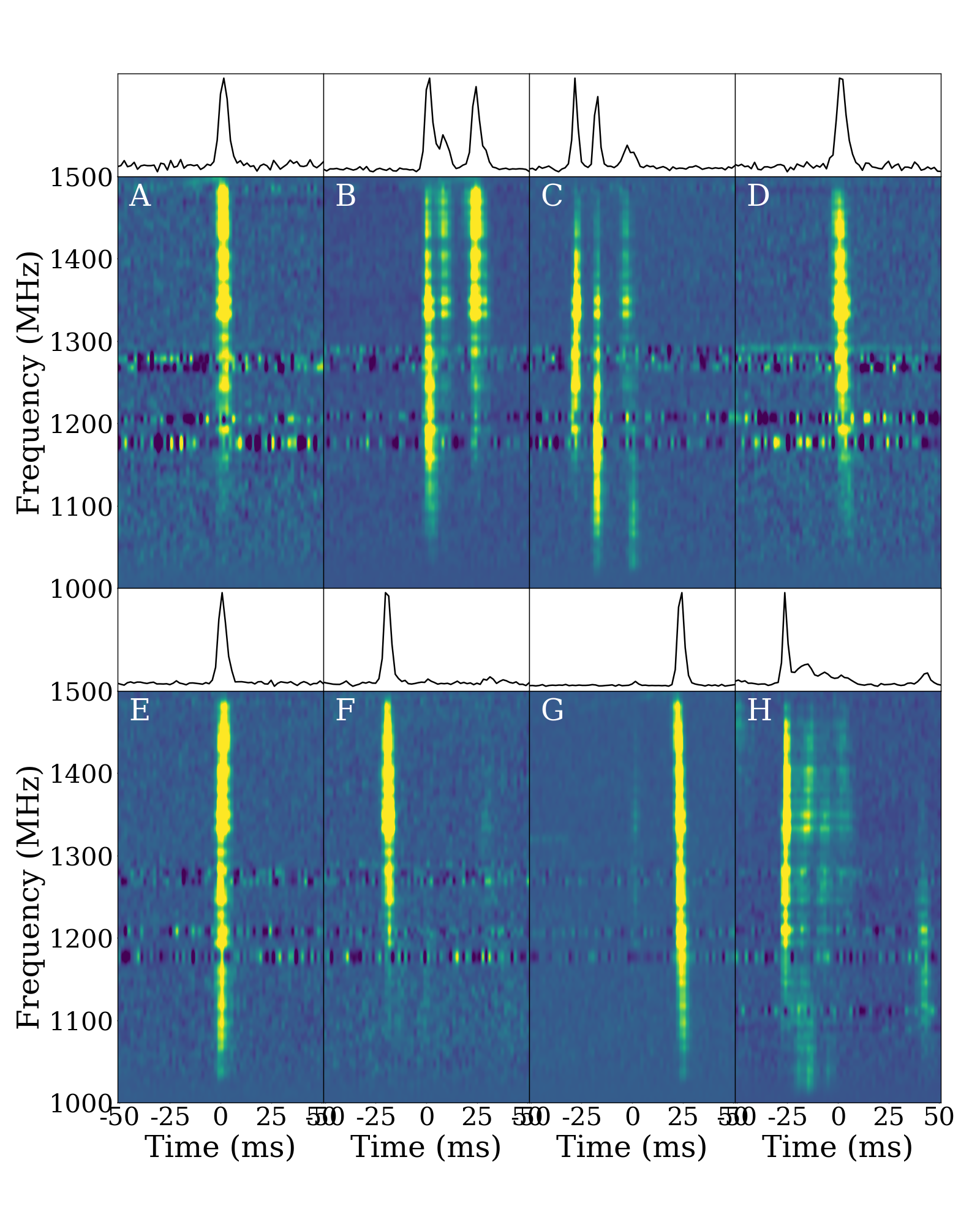}
    \caption{De-dispersed and de-scintillation dynamic spectra of eight golden sample bursts.}
    \label{fig:Observe ddm}
\end{figure}

With the solved images in 4096 channels, we can simulate a dynamic spectrum. We assume that the intrinsic temporary profile of FRB is Gaussian with the full width at half maximum (FWHM) of 1ms. The energy of FRB is assumed to be identical in each frequency channel. For the five cases in Figure \ref{fig:check_dynamic_spectra_2}, we show the corresponding mock dynamic spectra in Figure \ref{fig:plot_dynamic_spectra}. The dynamic spectra of eight golden sample bursts are shown in Figure \ref{fig:Observe ddm}. By comparison there are some requirements for the value of $\left. a^2 \small / D_{\mathrm{LS}}\right.$. Double caustic frequency peaks can be seen in the spectrum, and $\left. a^2 \small / D_{\mathrm{LS}}\right.$ is required to be lower than the values in (b); There is no separate components in the dynamic spectrum and no tail shape in the light curve requiring it to be lower than the values in (d)\footnote{A,D,E,F,G in Figure \ref{fig:Observe ddm} show single component in the dynamic spectrum and the light curve without a tail shape. We therefore suggest multiple components in B,C,H belonging to adjacent bursts.}. The constrain to $N_0$ is then given by Eq.(\ref{eq:the combined form of lens parameters}). Therefore, we suggest that if the double peaks in FRB 20201124A is caused by plasma lensing, for a typical filament structure with the characteristic width of 10AU at distance $\sim 1\mathrm{kpc}$ from the FRB, its maximum electron density $N_0$ is about $0.1\; \mathrm{pc\;cm^{-3}}$, and the transverse velocity is about 980 km/s; For the width of 1AU, $N_0$ is about $0.001\; \mathrm{pc\;cm^{-3}}$, and the transverse velocity is about 98 km/s.


\section{Summary}
\label{sec:Summary}
In this paper we suggest that the quenching phenomena after a sharp increase of burst counts in FRB 20201124A can be attributed to plasma lensing. Considering a plasma structure in the host galaxy, we define a lens property parameter $\mathrm{P}_0$ to combine the height $N_0$ and the width $a$ of the lens and its distance $D_{\mathrm{LS}}$ to the source as $\mathrm{P}_0 \propto \left ( \frac{a}{\mathrm{AU}}\right )\sqrt{\frac{\mathrm{kpc}}{D_{\mathrm{LS}}} \frac{\mathrm{pc}\;\mathrm{cm}^{-3}}{N_0} }$. Supposing a one-dimension Gaussian model, we calculate the scaled frequencies $\mathrm{P}_0 \nu$ ($\nu$ in GHz) corresponding to caustics for a given source position. We find that for a source position exceeding the minimum $y_{\mathrm{min}}=1.837$, there are two frequencies on caustic. To match the theory prediction we find that source position $y$ can be determined with the ratio of caustic frequencies. We suggest that if the sudden quenching of FRB 20201124A is due to plasma lensing, double main peaks can be found in spectra of bursts just before quenching.

We review the last four days' data from FAST observation before FRB 20201124A quenched in 2021. Data reduction procedures including de-dispersion and de-scintillation are performed to obtain spectra. We define a criterion to identify a main peak and found eight bursts appearing only two main-peaks at stable positions, i.e., golden sample. 

We estimate source positions according to the ratio of high frequency to low frequency of the eight golden sample bursts using one dimension Gaussian model. With theoretical source positions $y$ and according caustic frequencies $\mathrm{P}_0\nu$, we estimate the lens parameter $\mathrm{P}_0$ for each burst. They share almost the same values of about 0.6793 for all the golden sample bursts. We therefore obtain the combined lens parameters as $\left ( \frac{a}{\mathrm{AU}}\right )\sqrt{\frac{\mathrm{kpc}}{D_{\mathrm{LS}}} \frac{\mathrm{pc}\;\mathrm{cm}^{-3}}{N_0} } \approx 28.118$.  

We perform linear-fitting to theoretical source positions and bursts times for five bursts with monotonically decreasing positions and get a relative transverse motion velocity of $v\approx98\left(\frac{a}{\mathrm{AU}}\right)\mathrm{km/s}$. By comparing with possible quench time, we suggest that $v$ is relative reasonable. The scaled frequency range for FAST is estimated by $\mathrm{P}_0$. We therefore suggest that caustic frequency is still in the range of FAST observable window for the other three days before quenching. 
The reason for absence of significant stable double main peaks on these days may be FRB's inherent radiation mechanism. 

Finally, we solve lens equations through 4096 frequency channels for an average source position $y=1.88112$. We solved individual images in each frequency channel and calculate the magnification and arrival time. We find the two pairs of counter-images with high magnification images in two caustic frequency channels. Their maximum time delay $\delta t_{14}$ is modulated by $\left. a^2 \small / D_{\mathrm{LS}}\right.$, which affects the observed dynamic spectrum. By comparing with the dynamic spectra of the golden sample bursts, we suggest that $\left. a^2 \small / D_{\mathrm{LS}}\right.$ is less than $\left. \left(40\mathrm{AU}\right)^2 \small / c \cdot 1\mathrm{kpc}\right.$. Then the maximum electron density $N_0$ is constrained to be less than about $2 \, \mathrm{pc \; cm^{-3}}$ according to the value of $\left ( \frac{a}{\mathrm{AU}}\right )\sqrt{\frac{\mathrm{kpc}}{D_{\mathrm{LS}}} \frac{\mathrm{pc}\;\mathrm{cm}^{-3}}{N_0} }$. For a typical filament structure with the width of 1AU in host galaxy, e.g., $D_{\mathrm{LS}}\approx 1\mathrm{kpc}$, $N_0$ is about $0.001 \, \mathrm{pc \; cm^{-3}}$ and the transverse velocity is about $98\,\mathrm{km/s}$.

Due to some limitations, such as the lack of follow-up observation, we are still unable to provide a definitive conclusion on whether the suddenly quenching of FRB 20201124A at the end of September 2021 is caused by a plasma lens yet. Conclusions drawn in this paper are reasonable but relatively loose. Meanwhile we face a similar question as \citet{Levkov2022}: why were there only eight bursts with significant double main peaks on the first day before quenching? We think it may be due to the inherent radiation mechanism of FRB. However, in this paper, we provide a feasible explanation for this phenomenon. Taking observation data of FRB 20201124A and one-dimensional Gaussian model as an example, we are able to explain some observable features. And via these features, we can make limitations or constraints on the properties of plasma lens. Through the exploratory attempts in this work, we hope to propose a method for exploring the nature of FRB from its neighboring plasma environment. Some inconsistencies between data and theory demands a more accurate modeling of plasma lens, which is considered as our further works.

\section*{Acknowledgements}
We thank Ziwei Wu for helpful discussions and suggestions. Xuechun Chen acknowledges the support from Project funded by China Postdoctoral Science Foundation (No. 2023M730298). Di Li acknowledges the support from National Natural Science Foundation of China Grants No. 11988101. Pei Wang acknowledges support from the National Natural Science Foundation of China under grant U2031117, the Youth Innovation Promotion Association CAS (id. 2021055), and the Cultivation Project for FAST Scientific Payoff and Research Achievement of CAMS-CAS. Wenwen Zheng acknowledges the support from the CAS Project for Young Scientists in Basic Research (YSBR-063). The authors also thank the help from cosmology simulation database (CSD) in the National Basic Science Data Centre (NBSDC) and its funds the NBSDC-DB-10 (No. 2020000088).



\bibliographystyle{mnras}
\bibliography{manuscript}  





\bsp	
\label{lastpage}
\end{document}